\documentclass[runningheads]{llncs}
\usepackage{graphicx}

\usepackage{tikz}
\usepackage{comment}
\usepackage{amsmath,amssymb} 
\usepackage{color}

\begin{document}
\pagestyle{headings}
\mainmatter
\def\ECCVSubNumber{36}  


\title{Single image dehazing for a variety of haze scenarios using back projected pyramid network}

\titlerunning{BPPNet: a versatile back projected pyramid network for dehazing}
%
\author{{Ayush Singh\inst{1}} \and
Ajay Bhave\inst{1} \and
Dilip K. Prasad\inst{2}}
\authorrunning{A. Singh et al.}
%
\institute{Indian Institute of Technology (ISM), Dhanbad, India 826004, \email{ayush.s.18je0204@cse.iitism.ac.in} \and
UiT The Arctic University of Norway, Tromsø, Norway 9019,
\email{dilip.prasad@uit.no}\\
}
\maketitle

\begin{abstract}
Learning to dehaze single hazy images, especially using a small training dataset is quite challenging. We propose a novel generative adversarial network architecture for this problem, namely back projected pyramid network (BPPNet), that gives good performance for a variety of challenging haze conditions, including dense haze and inhomogeneous haze. Our architecture incorporates learning of multiple levels of complexities while retaining spatial context through iterative blocks of UNets and structural information of multiple scales through a novel pyramidal convolution block. These blocks together for the generator and are amenable to learning through back projection. We have shown that our network can be trained without over-fitting using as few as 20 image pairs of hazy and non-hazy images. We report the state of the art performances on NTIRE 2018 homogeneous haze datasets for indoor and outdoor images, NTIRE 2019 denseHaze dataset, and NTIRE 2020 non-homogeneous haze dataset.

\keywords{Single image dehazing, Generative adversarial network, Back projection, Deep learning}
\end{abstract}

\section{Introduction}

The quality of images of scenes in our daily life is greatly affected by the particles suspended in the environment, such as due to dust, smoke, mist, fog, smog, etc. Bad weather also contributes to this. Beside significantly higher and non-uniform noise in the images, the usual effects are reduced visibility, reduced sharpness, and contrast of the objects within the visibility and obscuring of other objects. Therefore, performing computer vision tasks like object detection, object recognition, tracking and segmentation becomes complicated for such images. Therefore, the true potential of computer vision empowered automated and remote surveillance systems such as drones and robots cannot be realized under hazy conditions. Thus, it is of interest to enhance the quality of images taken under homogeneous and non-homogeneous hazy conditions and recover the details of the scene. Haze removal or dehazing algorithms address this problem.

\begin{figure}[t]
\centering
\includegraphics[width=0.7\linewidth]{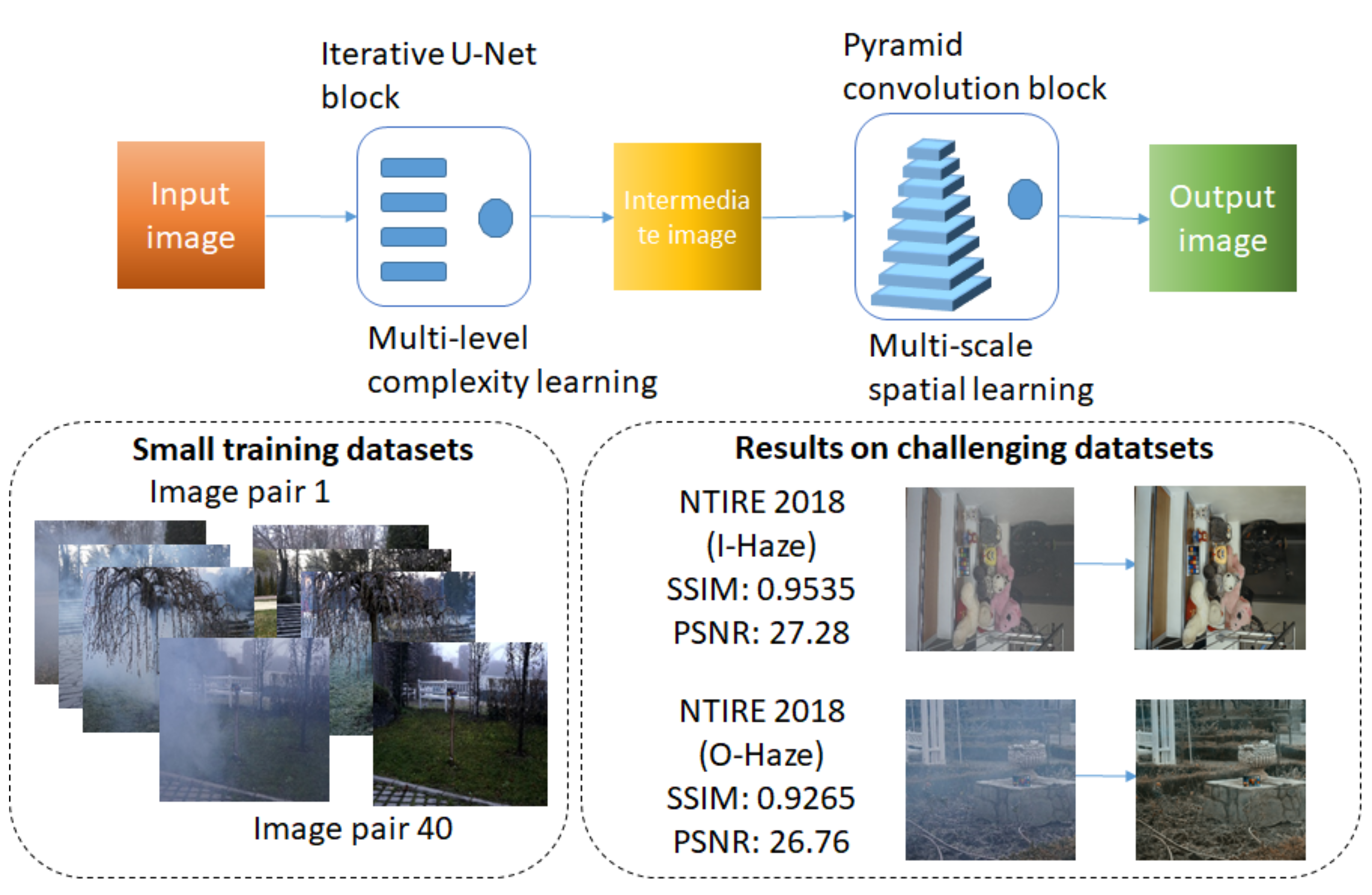}
\vspace{-3pt}
\caption{A compact representation of our novel generator and its important features.}
\label{img:UNet}
\end{figure}

There has been a significant activity in the topic of dehazing in recent years. New algorithms ranging from physics-based solvers, image processing based algorithms, as well as deep learning-based approaches, are being proposed. Furthermore, newer challenges are being undertaken, including dehazing in the presence of dense haze, non-homogenous haze, and using a single RGB image of a scene. It is being recognized that deep learning architectures provide better performance than the other approaches for diverse and challenging dehazing scenarios if suitably designed large datasets are available. However, dehazing images through deep learning on a small dataset using a single RGB image is quite challenging and of significant practical interest. For example, in the situation of fire management or natural disaster management, a suitable dehazing model characteristic of the situation needs to be learned quickly using a small number of images in haze and corresponding pre-disaster images.

We propose a novel deep learning architecture that is amenable to reliable learning of dehazing model using a small dataset. Our novel generative adversarial network (GAN) architecture includes iterative blocks of UNets to model haze features of different complexities and a pyramid convolution block to preserve and restore spatial features of different scales. The key contributions of this paper are as follows: 
\begin{itemize}
    \item A novel technique named pyramid convolution is introduced for dehazing to obtain spatial features of multiple scales structural information.
    \item We have used iterative UNet block for image dehazing tasks to make the generator learn different and complex features of haze without the loss of local and global structural information or without making the network too deep to result into loss of spatial features.
    \item The model used is end-to-end trainable with hazy image as input and haze-free image as the desired output. Therefore the conventional practice of using the atmospheric scattering model is obviated, and the problems encountered in inverse reconstruction are circumvented. It also makes the approach more versatile and applicable to haze scenarios where the conventional simplified atmospheric model may not apply.   
    \item Extensive experimentation is done on four contemporary challenging datasets, namely I-Haze and O-Haze datasets of NTIRE 2018 challenge, Dense-haze dataset of NTIRE 2019 challenge, and non-homogeneous dehazing dataset of NTIRE 2020 challenge. 
\end{itemize}

The outline of the paper is as follows. Section \ref{sec:related} presents related work, and section \ref{sec:proposed} introduces our architecture and learning approach. Section \ref{sec:results} presents numerical experiments and results. Section \ref{sec:ablation} includes an ablation study on the proposed method. Section \ref{sec:conclusion} concludes the paper.

\section{Related work} \label{sec:related}

Since this paper's focus is single image dehazing, we exclude a discussion on studies that required multiple images, for example, exploiting polarization, to perform dehazing. Single image dehazing is an ill posed problem because the number of measurements is not sufficient for learning the haze model, and the non-linearity of the haze model implies higher sensitivity to noise. Single image based dehazing exploits polarization-independent atmospheric scattering model proposed by Koschmieder \cite{koschmieder1925theorie} and its characteristics such as dark channel, color attenuation and haze-free priors. According to this model, the hazy image is specified by the atmospheric light (generally assumed uniform), the albedo of the objects in the scene, and the transmission map of the hazy medium. More details can be found in \cite{koschmieder1925theorie} and its subsequent citations,including recent ones \cite{chen2019multi,vazquez2020physical}. We have to predict the unknown transmission map and  global atmospheric light. In the past, many methods have been proposed for this task. The methods can be divided into two categories, namely (i) Traditional handcrafted prior based methods and (ii) Learning based methods. 

\textbf{Traditional handcrafted prior based methods:} Fattal~\cite{fattal2008single} proposed a physically grounded method by estimating the albedo of the scene. Tan ~\cite{tan2008visibility} proposed the use of the Markov random field to maximize the local contrast of the image. 
He et al.~\cite{he2010single} proposed a dark channel prior for the estimation of the transmission map. Fattal ~\cite{fattal3dehazing} proposed a color-line method based on the observation that small image patches typically exhibit a one-dimensional distribution in the RGB color space. Traditional handcrafted prior methods give good results for certain cases but are not robust for all the cases.

\textbf{Learning based approaches:} In recent years, many learning based methods have been proposed for single image dehazing that encash the success of deep learning in image processing tasks, availability of large datasets, and better computation resources. Some examples are briefly mentioned here. Cai et al.~\cite{cai2016dehazenet} proposed an end-to-end CNN based deep architecture to estimate the transmission map. Ren et al.~\cite{ren2016single} proposed a multi-scale deep architecture, which also estimates the transmission map from the haze image. Zhang et.al. in~\cite{zhang2018densely} proposed a deep network architecture that estimates the transmission map and the atmospheric light. These estimates are then used together with the atmospheric scattering model to generate the haze-free image. 

\textbf{Our approach in context: }In contrast to these approaches, our approach is an end-to-end learning based approach in which the learnt model directly predicts the haze-free image without needing to reconstruct the transmission map and the atmospheric light, or using the atmospheric scattering model. It is therefore more versatile to be trained for situations where the atmospheric scattering model of \cite{koschmieder1925theorie} may not apply or may be too simple. Example includes non-uniform haze. It also circumvents the numerical errors and artifacts associated with the use of inverse approaches of reconstructing the haze-free image from the transmission map and atmospheric light.

\section{Proposed method}\label{sec:proposed}

In this section we present our model, namely back projected pyramid network (BPPNet). The overall architecture is based on generative adversarial network ~\cite{goodfellow2014generative}, where a generator generates a haze-free image from a hazy image, and a discriminator tells whether the image provided to it is real or not. 

\subsection{Generator} 
\begin{figure*}[t]
\includegraphics[width=\linewidth]{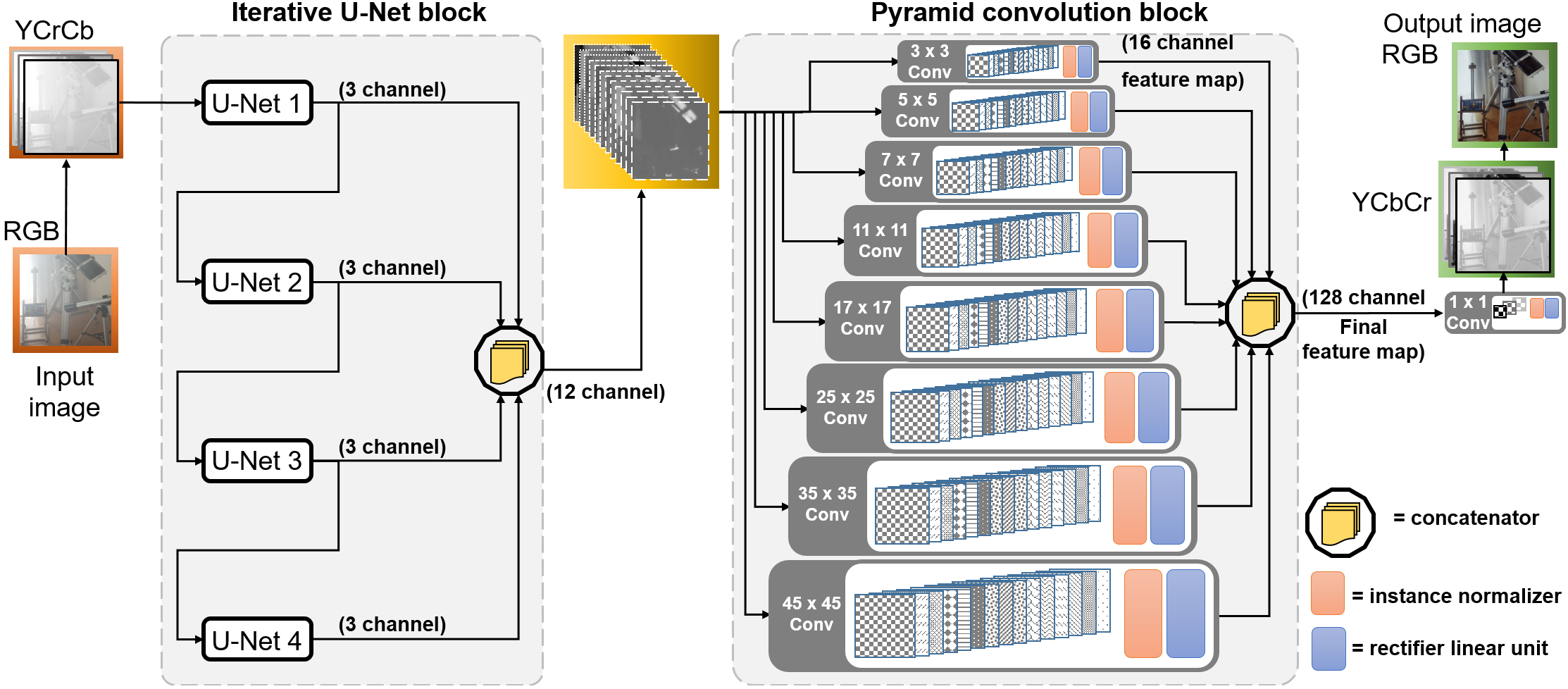}
\caption{The architecture of our generator.}
\label{img:generator}
\end{figure*}

The architecture of the generator is shown in Fig. \ref{img:generator}. It comprises of two blocks in series, namely (i) iterative UNet block, (ii) pyramid convolution block, which we describe next.

{\textbf{Iterative UNet block (IUB):}} This block consists of multiple UNet~\cite{ronneberger2015u} units connected in series i.e. the output of one UNet (architecture in the supplementary) is fed as the input to the next UNet. In addition, the output of each UNet is passed to a concatenator, which concatenates the 3 channel output of all the UNets, providing an intermediate 12 channel feature map. The equations describing the working of IUB are the following.

\begin{equation}
    I_{_1} = {\rm UNET}_1( I_{\rm haze} ); I_{{i}} = {\rm UNET}_{i}( I_{{i-1}}) \quad {\rm{for}} \quad i>1,
\end{equation}

\noindent where $I_i$ is the output of $i$th UNet unit, $I_{\rm haze}$ is the input hazy image after being transformed to YCbCr space, and the output $\hat{I}_{\rm IUB}$ of IUB is given as 
\begin{equation}
    \hat{I}_{\rm IUB} = I_{1}\oplus I_{2}\oplus \ldots I_{M}
\end{equation}
where $\oplus$ indicates concatenation operator and $M$ is the number of UNet unit. We have used $M=4$. An ablation study on the value of $M$ is presented later in section \ref{sec:ablation}. Here, we discuss the need of more than one UNet.

In principle, a single UNet may be able to support dehazing to some extent. However, it may not be able to extract complex features and generate an output with fine details. One way to tackle this problem is to increase the number of layers in the encoder block so that more complex features can be learned. But the layers in the encoder block are arranged in feed forward fashion, and the height and the width of layers decreases upon moving further. This causes loss in spatial information and reduces the possibility of extraction of spatial features of high complexity. Therefore, we take an alternate approach of creating sequence of the multiple UNets. The sequence of UNets may be interpreted as a sequence of multiple encoder-decoder pairs with skip connections. The encoder in each UNet extracts the features from input tensor in the downsample feature map and decoder uses those features and projects them into an upsample latent space with same height and width as input tensor. In this way, each generator helps in learning increasingly complex features of haze while the decoder helps in retaining the spatial information in the image. Lastly, the concatenation step ensures that complexity of all the levels are available for subsequent reconstruction.

\begin{figure*}[t]
\centering
\includegraphics[width=\linewidth]{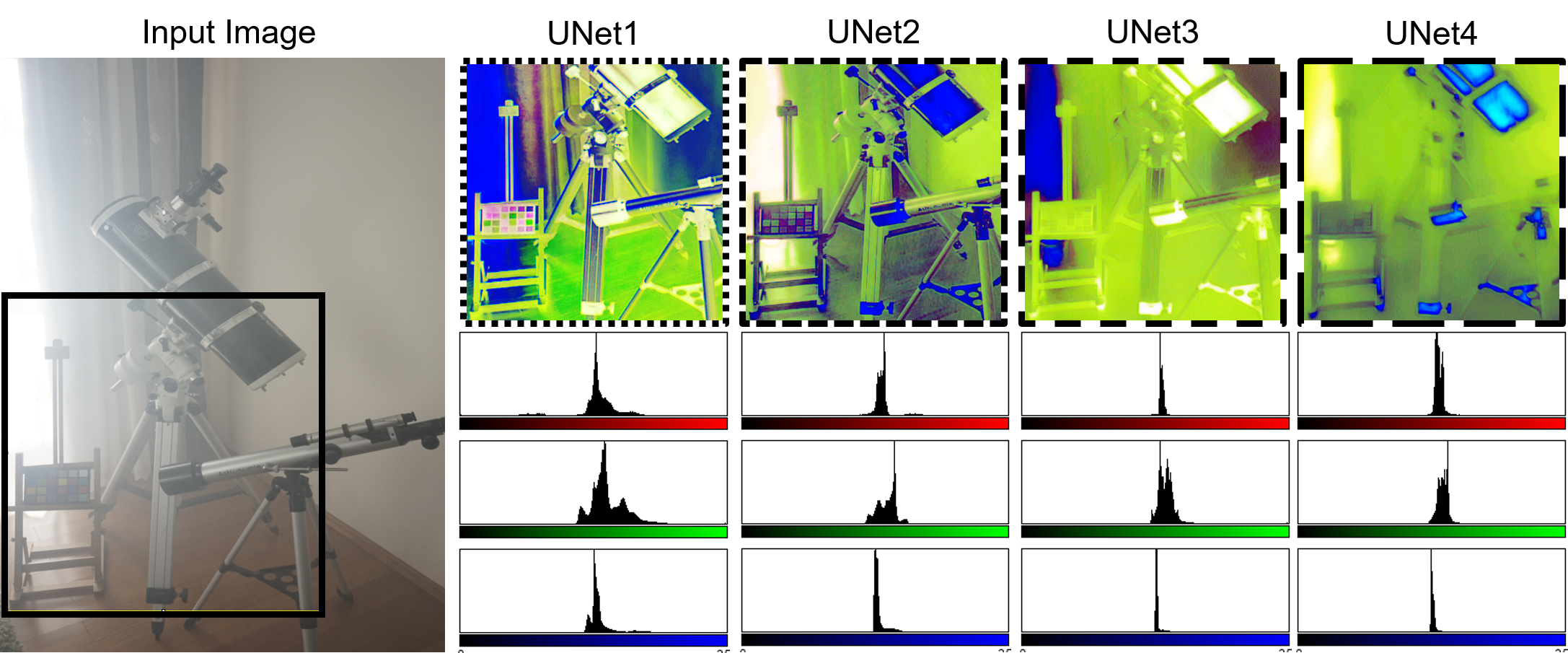}
\caption{The effect of the successive UNet units is illustrated. Images are histogram equalized for better visualization. The histograms of the channels becomes narrower after passing more number of UNet units, indicating that adding more UNet units may cease to create more value after a certain limit.}
\label{img:iedb}
\end{figure*}

We illustrate the effect of using multiple UNets in Fig. \ref{img:iedb}. Histogram equalized 3-channel output of each UNet is shown as an RGB image. It is seen that the spatial context is preserved, and at the same time haze introduced blur of different complexities are present in the outputs of different images. The haze in the last UNet output is flatter across the image and shows large scale blurs while the haze in the first UNet is local and introduces small scale blurs. Therefore, most dehazing is accomplished in UNet1, although the subsequent UNets pick the dehazing components that the previous UNets did not. Fig. \ref{img:iedb} also explains our choice of only four UNet blocks even though more blocks could be used in principle. We explain our choice in two parts. First, there is a trade-off involved between accuracy and speed when choosing the number of UNet blocks. Second, as seen in the histograms in Fig. \ref{img:iedb}, the dynamic range of channels decreases with every subsequent UNet block, thereby indicating the reduction in the usable information content. The standard deviation of the intensity values in the 3 channels after UNet4 is $\sim$12.2. Adding more blocks would further reduce this value, and therefore not provide significantly exploitable data for dehazing. 

\begin{figure*}[t]
\centering
\includegraphics[width=0.6\linewidth]{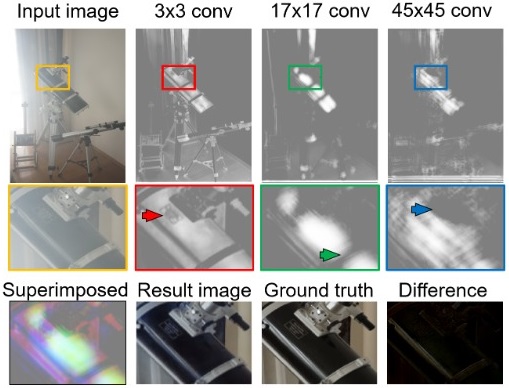}
\caption{Feature maps corresponding to one of the channels of 3$\times$3, 17$\times$17, and 45$\times$45 convolution layer respectively. The figure shows that smaller kernel size generates smaller scale features such as edges while large kernel size generates large scale features such as big patches.}
\label{img:newfig}
\end{figure*}

{\textbf{Pyramid convolution (PyCon) block:}} 
Although the iterative UNet block does provide global and local structural information, the output lacks the global and local structural information for different sized objects. An underlying reason is that the structural information from different scales are not directly used to generate the output.  To overcome this issue, we have used a novel pyramid convolution technique. Earlier pyramid pooling has been used in~\cite{ronneberger2015u} to leverage the “global structural information”. However, since the pooling layers are not learnable, we instead employ learnable convolution layers that can easily outperform the pooling layers in leveraging the information. 

We employ many convolution layers of different kernel sizes in parallel on the input map (the 12 channel output of iterative UNet block). Corresponding to different kernel sizes used for convolution, different output maps are generated with structural information of different spatial scales. The kernel sizes are chosen as 3, 5, 7, 11, 17, 25, 35, and 45, as shown in Fig. \ref{img:generator}. Odd sized kernels are used since pixels in the intermediate feature map are assumed to be symmetrical around output pixel. We observed introduction of distortion over layers upon using even-sized kernels, indicating the importance of exploiting the symmetry of the features around the output pixel. Zero padding is used to ensure that the features at the edges are not lost. After the generation of feature map from corresponding kernels, all the generated maps are concatenated to make an output feature map of 128 channels, which is subsequently used for the final image construction by applying a convolution layer of kernel size 3$\times$3 with zero padding.  In this manner, the local to global information is directly used for the final image reconstruction.

The effect of using pyramid convolution is shown in Fig. \ref{img:newfig}. In the zoom-ins shown in the middle panel, the arrows indicate some features of the size of the convolution filter used for generating that particular feature map. The illustrated 3 channels are superimposed as a hypothetical RGB image in the bottom left of Fig. \ref{img:newfig} to demonstrate the types of details present in a subset of the output feature map. Since we have used 8 convolution filters that operate on 12 channel input, we generate a total of 128 channels in the output feature map with a large variety of spatial features of multiple scales learned and restored. Therefore, the result image shown in the bottom panel has spatial features closely matching the ground truth, resulting in a low difference map (shown in the bottom right). 

One may consider using the 12 channel output of the iterative UNet block for generating the dehazed image directly, without employing the PyCon block. To indicate the importance of including the PyCon block, we include an ablation study in section \ref{sec:ablation}. 

\begin{figure}[t]
    \centering
    \includegraphics[width=0.75\linewidth]{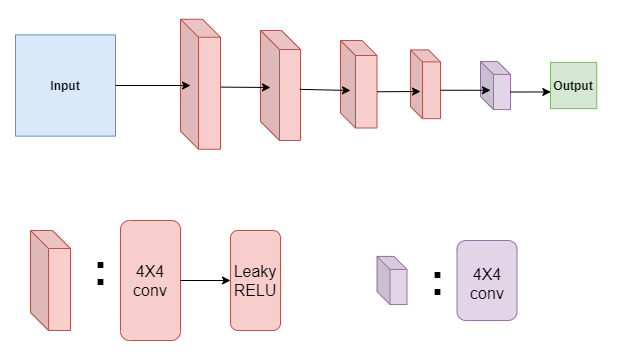}
    \caption{Architecture of the discriminator.}
    \label{fig:discriminator}
\end{figure}

\subsection{Discriminator} We have used a patch discriminator to determine whether a particular patch is realistic or not. The patches overlap in order to eliminate the problem of low performance on the edges. We have used 4$\times$4 convolution layers in discriminator. After every convolution layer, we have added an activation layer with an activation leaky rectified linear unit (Leaky ReLu) except the last layer where the activation function is sigmoid. The size of the convolution kernel used is 4$\times$4, and the output map size is 62$\times$62 for an input of size 512$\times$512. The architecture of our discriminator is shown in Fig. \ref{fig:discriminator}. 

\subsection{Loss functions}\label{sec:losses}
Most dehazing models use the mean squares error (MSE) as the loss function \cite{zhang2018multi}. However, MSE is known to be only weakly correlated with human perception of image quality \cite{huang2014visibility}. Hence, we employ additional loss functions that are closer to human perception. To this end, we have used a combination of MSE ($L_2$ loss), adversarial loss $L_{\rm{adv}}$, content loss $L_{\rm{con}}$, and structural similarity loss $L_{\rm{SSIM}}$. We define the remaining loss functions below. 

The adversarial loss for the generator $L_{\rm{adv}}$ and the discriminator $L_{\rm dis}$ is defined as:

\begin{equation}
    L_{\rm adv} = \langle \log(D(I_{\rm pred})) \rangle,
\end{equation}
\begin{equation}
    L_{\rm dis} = \langle \log(D(I_{\rm GT})) \rangle + \langle \log(1-D(I_{\rm pred})) \rangle,
\end{equation}

\noindent where $(I_{\rm haze}, I_{\rm GT})$ are the supervised pair of the hazy image and the corresponding ground truth. 
$D(I)$ is the discriminator’s estimate of the probability that data instance $I$ provided to it is indeed real. Similarly, $G(I)$ is the generator output for the input instance $I$. Further, $I_{\rm pred} = G (I_{\rm haze})$. The notation $<>$ indicates the expected value over all the supervised pairs. 

The MSE, also referred to as the $L_2$ loss, is defined as the average norm 2 distance between $I_{\rm{GT}}$ and $I_{\rm{pred}}$:

\begin{equation}
    L_2= \langle{I_{\rm{GT}}-I_{\rm{pred}}}\rangle
    \label{eq:L2}
\end{equation}

Our content loss is the VGG based perceptual loss~\cite{johnson2016perceptual}, defined as:

\begin{equation}
    L_{\rm{con}}= \big\langle\sum_i{\frac{1}{N_i}{\mid\mid \phi_i(I_{\rm GT})-\phi_i(I_{\rm pred})\mid\mid}}\big\rangle,
\end{equation}

\noindent where $N_i$ is the number of elements in the $i^{th}$ activation of VGG-19 and $\phi_i$ represents $i^{\rm{th}}$ activation of VGG-19. 

The structural similarity loss $L_{\rm{SSIM}}$ over reconstructed image $I_{\rm pred}$ and ground truth $I_{\rm GT}$ is defined as:

\begin{equation}
    L_{\rm{SSIM}}= 1 - \langle {{\rm SSIM}(I_{\rm GT},I_{\rm pred})}\rangle,
\end{equation}

\noindent where ${\rm SSIM}(I, I^\prime)$ is the SSIM \cite{wang2004image} between the images $I$ and $I^\prime$. We note that although the losses $L_2$ and $L_{\rm SSIM}$ directly compare the predicted and the ground truth images, the nature of comparison is quite different across them. $L_2$ is insensitive to the structural details but retains the comparison of the general energy and dynamic range of the two images being compared. $L_{\rm SSIM}$ on the other hand compared the structural content in the images with less sensitivity to the contrast. Therefore, including these two loss functions provide complementary aspects of comparison between the predicted and the ground truth images.   

The overall generator loss $L_{\rm{G}}$ and discriminator loss $L_{\rm{D}}$ are given as 

\begin{equation}
    {L_{\rm{G}}=A_1 L_{\rm{{adv}}}+A_2 L_{\rm{con}}+A_3 L_2 + A_4 L_{\rm{SSIM}}} 
\end{equation}
\begin{equation}
{L_{\rm{D}}=B_1 L_{\rm{D_{adv}}}.}
\end{equation}

We have heuristically chosen the values of the constant weights in the above equation as $A_1 = 0.7$, $A_2 = 0.5$, $A_3 = 1.0$, $A_4 = 1.0$, and $B_1 = 1.0$.

\section{Experimental results}\label{sec:results}

\subsection{Datasets}
We have trained and tested our model on the following four datasets, namely NTIRE 2018 image dehazing indoor dataset (abbreviated as I-Haze), NTIRE 2018 image dehazing outdoor dataset (O-Haze), Dense-Haze dataset of NTIRE 2019, and NTIRE 2020 dataset for non-homogeneous dehazing challenge (NTIRE 2020).

\textbf{\textit{I-Haze ~\cite{DBLP:journals/corr/abs-1804-05091} and O-Haze ~\cite{ancuti2018haze}:}} These datasets contains 25 and 35 hazy images (size 2833$\times$4657 pixels) respectively for training. Both datasets contain 5 hazy images for validation along with their corresponding ground truth images. For both of these datasets, the training was done on training data and validation images were used for testing because although 5 hazy images for testing were given but their ground truths were not available to make the quantitative comparison. 

\textbf{\textit{Dense-Haze ~\cite{ancuti2019dense}:}} This dataset contains 45 hazy images (size 1200$\times$1600 pixels) for training and 5 hazy images for validation and 5 more for testing with their corresponding ground truth images. We have done training on training data and tested our model with test data.

\textbf{\textit{NTIRE 2020~\cite{NTIRE2020}:}} This dataset contains 45 hazy images (size 1200$\times$1600 pixels) for training with their corresponding ground truth. It is the first dataset of non-homogeneous haze in our knowledge. As ground truth for validation was not given, hence we used only 40 image pairs for training and calculated our results on the rest of the 5 images.

\begin{table}[t]
\centering
\caption{Quantitative comparison of various state of the art methods with our model on I-Haze, O-Haze and Dense-Haze datasets.Our model does the dehazing task in real-time at an average running time of 0.0311 s i.e. 31.1 ms per image.}
\label{tbl:indoor_outdoor_table}
\begin{tabular}{|l||c|c|c|c|c|c|c|c|}
    \hline
    \multicolumn{9}{|c|}{I-Haze dataset}\\
    \hline
    Metric & Input & CVPR'09 & TIP'15 & ECCV'16 & CVPR'16 & ICCV'17 & CVPRW'18 & Our\\
    & & \cite{he2010single}& \cite{zhu2015fast}& \cite{ren2016single}& \cite{berman2016non}& \cite{li2017all}& \cite{zhang2018multi}& model\\
    \hline
    SSIM  & 0.7302 & 0.7516 & 0.6065 & 0.7545 & 0.6537 & 0.7323 & 0.8705 & \textbf{0.8994}\\
    \hline
    PSNR & 13.80 & 14.43 & 12.24 & 15.22 & 14.12 & 13.98 & 22.53 & \textbf{22.56}\\
    \hline
\end{tabular}
\begin{tabular}{|l||c|c|c|c|c|c|c|c|}
    \hline
    \multicolumn{9}{|c|}{O-Haze dataset}\\
    \hline
    Metric & Input & CVPR'09 & TIP'15 & ECCV'16 & CVPR'16 & ICCV'17 & CVPRW'18 & Our\\
    & & \cite{he2010single}& \cite{zhu2015fast}& \cite{ren2016single}& \cite{berman2016non}& \cite{li2017all}& \cite{zhang2018multi}& model\\
    \hline
    SSIM  & 0.5907 & 0.6532 & 0.5965 & 0.6495 & 0.5849 & 0.5385 & 0.7205 & \textbf{0.8919}\\
    \hline
    PSNR & 13.56 & 16.78 & 16.08 & 17.56 & 15.98 & 15.03 & 24.24 & \textbf{24.27}\\
    \hline
\end{tabular}
\begin{tabular}{|l||c|c|c|c|c|c|c|c|c|}
    \hline
    \multicolumn{10}{|c|}{Dense-Haze dataset}\\
    \hline
    Metric & CVPR & Meng & Fattal & Cai & Ancuti & CVPR & ECCV & Morales & Our\\
    & '09 \cite{he2010single} & et. al \cite{meng2013efficient} & \cite{fattal3dehazing} & et. al \cite{cai2016dehazenet} & et. al \cite{ancuti2016night} & '16 \cite{berman2016non} & '16 \cite{ren2016single} & et. al \cite{morales2019feature} & model\\
    \hline
    SSIM  & 0.398 & 0.352 & 0.326 & 0.374 & 0.306 & 0.358 & 0.369 & 0.569 & \textbf{0.613}\\
    \hline
    PSNR & 14.56 & 14.62 & 12.11 & 11.36 & 13.67 & 13.18 & 12.52 & 16.37 & \textbf{17.01}\\
    \hline
\end{tabular}
\end{table}

\begin{figure}[t]
\centering
\includegraphics[width=\linewidth]{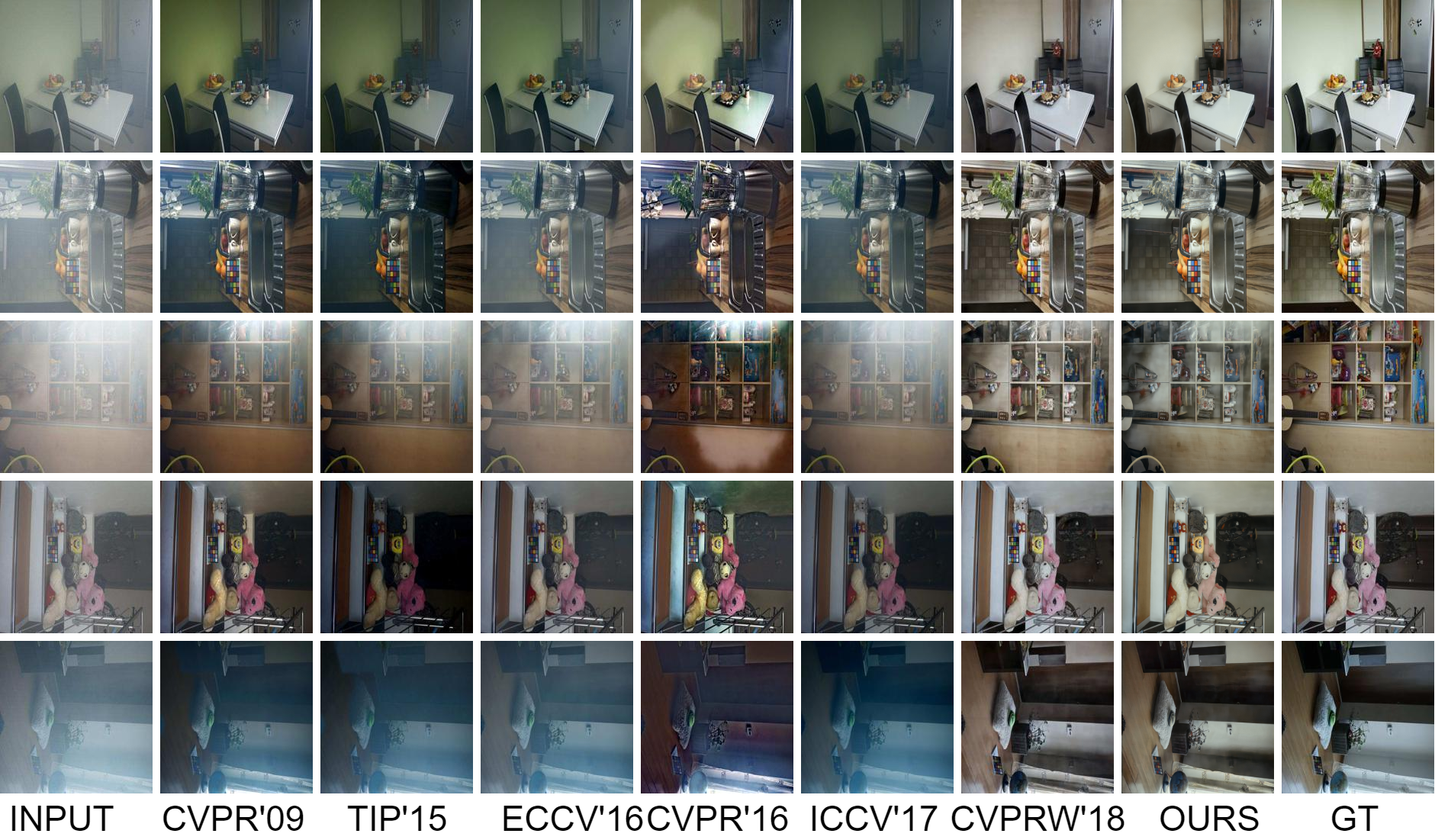}
\caption{Qualitative comparison of various benchmark with our model on I-Haze dataset}
\label{img:indoor_res}
\end{figure}
\begin{figure}[t]
\centering
\includegraphics[width=\linewidth]{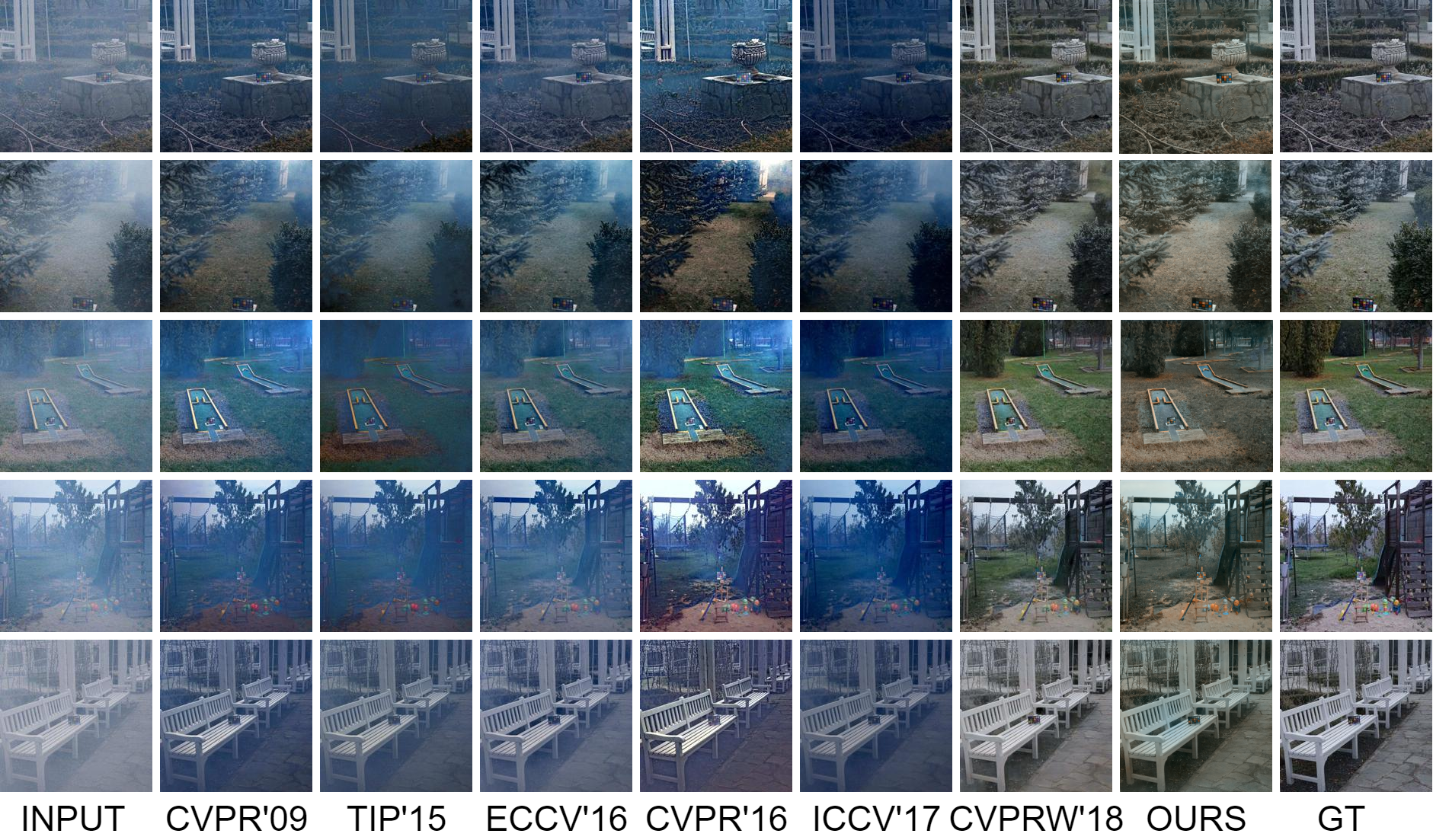}
\caption{Qualitative comparison of various methods with our model on O-Haze dataset.}
\label{img:outdoor_res}
\end{figure}
\begin{figure}[t]
    \centering
    \includegraphics[width=\linewidth]{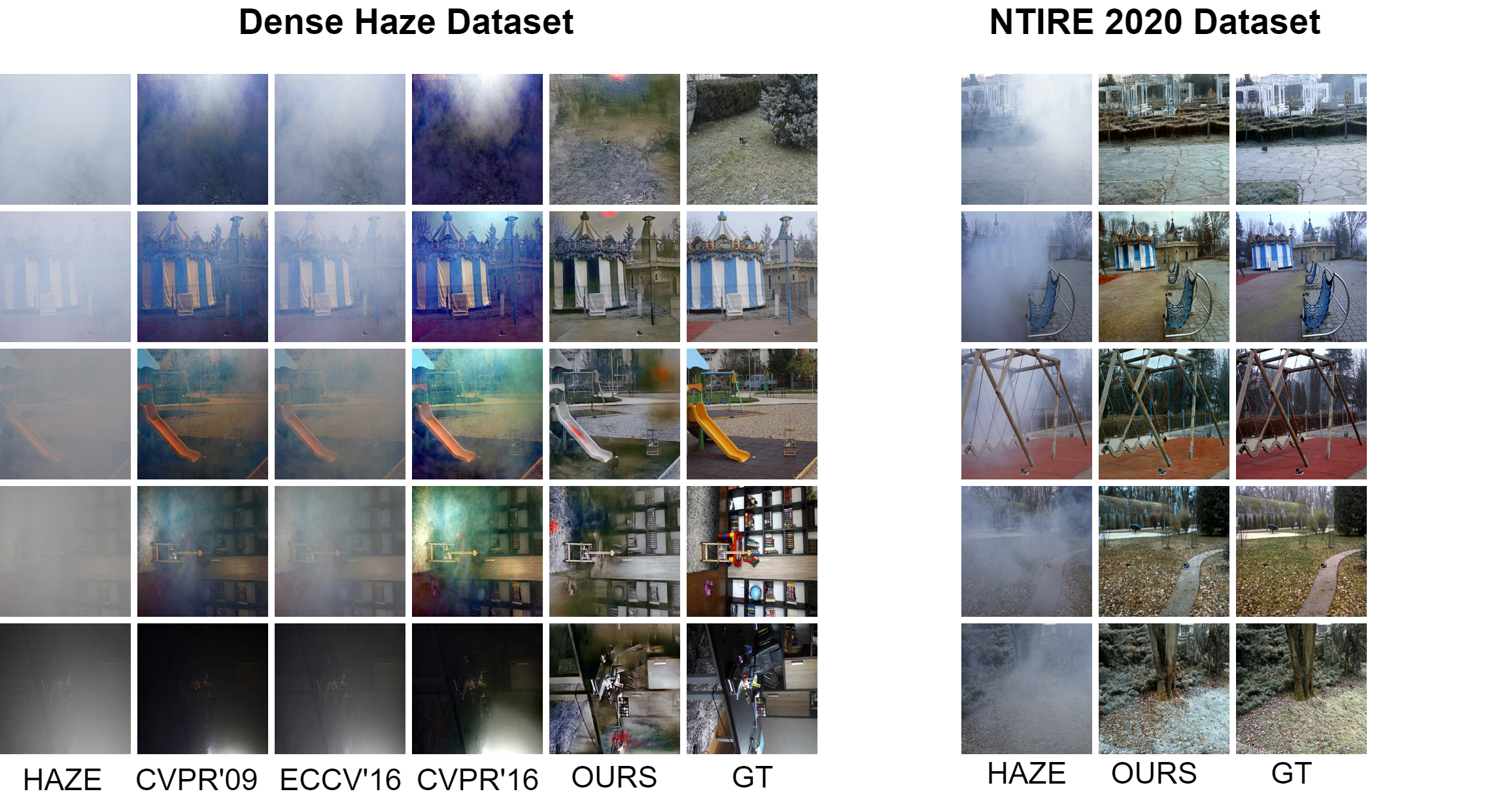}
    \caption{Qualitative results for Dense-Haze and NTIRE 2020 dataset.}
    \label{img:dense}
\end{figure}

\subsection{Training details}
The optimizer used for the training was Adam~\cite{kingma2014adam} with the initial learning rate for 0.001 and 0.001 for generator and discriminator respectively. We have randomly cropped large square patches from the training images. The crop size was 1024$\times$1024 for NTIRE 2020 and Dense-Haze. Leveraging on the even large sizes of images in I-Haze and O-Haze datasets, we created four crops each of two different sizes, 1024$\times$1024 and 2048$\times$2048. We then resized all the patches to 512$\times$512 using bicubic interpolation. These patches are randomly cropped for each epoch i.e. these patches are not same for every epoch. This has created an effectively larger dataset out of the small dataset available for training for each of the considered datasets. For datasets named I-Haze, O-Haze and NTIRE 2020, we converted these randomly cropped resize patches from RGB space to YCbCr space and then used them for training. For Dense-Haze dataset we directly used RGB patches for training. 

We decreased the learning rate of the generator whenever the loss became stable. We stopped training when the learning rate of the generator reached 0.00001 and the loss stabilized. We also tried to decrease the learning rate of discriminator but found that doing so did not give the best results.

\subsection{Results} 
Here, we present our results and their comparison with the results of other models available in the literature. We note that we converted the test input image size to 512$\times$512 for all the datasets for generating our results in view of our hardware (GPU) memory constraints. Quantitative evaluation is performed using the SSIM metric and the peak signal to noise ratio (PSNR). The metrics are computed in the RGB space even if the training was done in YCbCr space. The quantitative  results are compared with earlier state-of-the-art in Table \ref{tbl:indoor_outdoor_table}. The metrics for the other methods are reproduced from ~\cite{zhang2018multi} for the I-Haze and O-Haze dataset. The benchmark for Dense-haze was provided in ~\cite{ancuti2019dense}. We further include the results of Morales at al.~\cite{morales2019feature} for comparison.

\textbf{\textit{I-Haze:}} The average PSNR and SSIM of our method for this dataset on validation data were 22.56 and 0.8994 respectively. It is evident from  Table \ref{tbl:indoor_outdoor_table} that our model outperforms the state-of-the-art in both SSIM and PSNR index by a good margin. Qualitative comparison results on the test data are shown in Fig. \ref{img:indoor_res}. It is evident that only CVPRW'18 \cite{zhang2018multi} competes with our method in the quality of dehazed image and match with the ground truth. 

\textbf{\textit{O-Haze:}} The average PSNR and SSIM for this dataset on validation data were 24.27 and 0.8919 respectively on validation data, see Table \ref{tbl:indoor_outdoor_table}. Our model clearly outperforms all the state-of-the-art in both PSNR and SSIM index by a large margin. For SSIM, the closest performing method was CVPRW'18~\cite{zhang2018multi} with SSIM of 0.7205, which is significantly lower than ours i.e 0.8919. It is notable from the results of all the methods that this dataset is more challenging than I-Haze. Nonetheless, our method provides comparable performance over both I-Haze and O-Haze datasets. The qualitative comparison of results on the test data are shown in Fig. \ref{img:outdoor_res}. Similar to the I-Haze dataset, only CVPRW'18 \cite{zhang2018multi} and our method generate dehazed images of good quality. 

As compared to I-Haze results in Fig. \ref{img:indoor_res}, it is more strongly evident in Fig. \ref{img:outdoor_res} that the color cast of our method is slightly mismatched with the ground truth, where CVPRW'18 performs better than our method. However, CVPRW'18 shows poorer structural continuity than our method, as evident more strongly in Fig. \ref{img:indoor_res}.  

\textbf{\textit{Dense-Haze:}} From Table \ref{tbl:indoor_outdoor_table}, it is evident that this dataset is significantly more challenging that the I-Haze and O-Haze datasets. All methods perform quite poorer for this dataset, as compared to the numbers reported for I-Haze and O-Haze dataset. Even though the performance of our method is also poorer for this dataset as compared to the other datasets, its SSIM and PSNR values are significantly better than the other 8 methods whose results are included in Table \ref{tbl:indoor_outdoor_table} for this dataset. Qualitative comparison with select methods is shown in Fig. \ref{img:dense}. The results clearly illustrate the challenge of this dataset as the features and details of the scene are almost imperceptible through the dense haze. Only our method is capable of dehazing the image effectively and bringing forth the details of the scene. Nonetheless, the color cast is non-uniformly compensated and different from the ground truth in regions.  

\textbf{\textit{NTIRE 2020:}} As the ground truth for test data is not given, we randomly chose 5 images for testing and used the rest of the 40 image pair for training. The average SSIM and PSNR are 0.8726 and 19.40  respectively. This SSIM value is better than the best SSIM observed in the competition and informed to the participants in a personal email after the test phase. The qualitative results are shown in Fig. \ref{img:dense}. The observations are generally similar to the observations for the Dense-Haze dataset. Our results are qualitatively quite close to the ground truth and show the ability of our method to recover the details lost in haze, despite the non-homogeneity of the haze. Second, we observe a little bit of mismatch in the color reproduction and in-homogeneity in the color cast, which needs further work. We expect that the problem of color cast inhomogeneity may be related to the inhomogeneity in the haze itself, which may have been present in the Dense-Haze data as well but may not have been perceptible due to the generally high density of haze. 

\section{Ablation study}\label{sec:ablation}

We conduct ablation study using I-Haze and O-Haze datasets. We consider the ablation associated with the architectural elements in section \ref{sec:unet_experiment}, loss components in section \ref{sec:losses_experiment}, and the image space used in training in section \ref{sec:rgb_experiment}. 

\subsection{Architecture ablation} \label{sec:unet_experiment}

Here, we consider ablation study relating to the number of UNet units used in the iterative UNet block and the absence or presence of the pyramid convolution block. The results are shown in Table \ref{tab:abl1}. It is evident that decreasing or increasing the number of UNet blocks degrades the performance and the use of $M=4$ UNet blocks is optimal for the architecture. This is in agreement in the observations derived from Fig. \ref{img:iedb}. Similarly, dropping the PyCon block also degrades the performance. 

\subsection{Loss ablation} \label{sec:losses_experiment}
We proposed in section \ref{sec:losses} to use four types of loss functions for the training of the generator. Here, we consider the effect of dropping one loss function at a time. The results are presented in the bottom panel of Table \ref{tab:abl1}. It is seen than dropping any of the loss function results into significant deterioration of performance. This indicates the different and complementary roles each of these loss functions is playing. Our observation of the qualitative results, discussed in section \ref{sec:results}, we might need to introduce another loss function related to the preservation of the color cast or color constancy.

\begin{table}[t]
\centering
\caption{The results of ablation study are presented here. The reference indicates the use of 4 UNet blocks, inclusion of PyCon block with layer configuration as shown in Fig. \ref{img:generator}. All loss functions discussed in section \ref{sec:losses} are used and the entire architecture uses YCbCr space, such as shown in Fig. \ref{img:generator}.}
\label{tab:abl1}
\begin{tabular}{|l|l||c|c|c|c|c|c|}
\hline
\multicolumn{8}{|c|}{\textbf{(a) Ablation study on architectural units}}\\
\hline
{Dataset}& {Metric}&\textbf{Reference}&{1 UNet}&{2 UNets}&{3 UNets}&{5 UNets} &{No PyCon}\\
\hline
\hline
I-Haze & SSIM & \textbf{0.8994} & 0.8572 & 0.8679 & 0.8820 & {0.8932} & {0.8878}\\
\cline{2-8}
& PSNR & \textbf{22.57} & 18.54 & 19.94 & 20.92 & {21.62} & 21.17 \\
\hline
O-Haze & SSIM & \textbf{0.8927} & 0.8500 & 0.8639 & 0.8795 & {0.8639} & {0.8768}\\
\cline{2-8}
& PSNR & \textbf{24.30} & 21.34 & 22.36 & 23.06 & {22.36} & 23.13 \\
\hline
\hline
\multicolumn{8}{|c|}{\textbf {(b) Ablation study on losses and the image space}}\\
\hline
{}&{}&{Reference}&\multicolumn{4}{|c||}{The loss function dropped} & Direct use of RGB,\\
\cline{4-7}
{}&{}&{}&{$L_{\rm {adv}}$}&{$L_{\rm con}$}&{$L_{2}$}&{$L_{\rm SSIM}$} &{not the YCbCr space}\\
\hline
\hline
I-Haze & SSIM & \textbf{0.8994} & {0.8620} & {0.8372} & {0.8648} & {0.8343} & {0.8944}\\
\cline{2-8}
{} & PSNR & \textbf{22.57} & {19.52} & {18.99} & {20.02} & {19.58} & 20.94\\
\hline
O-Haze & SSIM & \textbf{0.8927} & {0.8608} & {0.8271} & {0.8650} & {0.8568} & {0.8712}\\
\cline{2-8}
{} & PSNR & \textbf{24.30} & {22.26} & {20.66} & {22.78} & {22.44} & 22.54\\
\hline
\end{tabular}
\end{table}

\subsection{Use of RGB versus YCbCr space} \label{sec:rgb_experiment}
If we used RGB space instead of YCbCr space for training, we observe a degraded performance in terms of SSIM as reported in section \ref{tab:abl1}(b). However, we note that this observation is not consistent over all the datasets. Specifically, we noted that for Dense-Haze, the YCbCr conversion gave little poorer results than RGB based training. Hence, we have used RGB patches for training on Dense-Haze dataset. 

\section{Conclusion}\label{sec:conclusion}
The presented single image dehazing method is an end-to-end trainable architecture that is applicable in diverse situations such as indoor, outdoor, dense, and non-homogeneous haze even though training datasets used are small in each of these cases. It beats the state-of-the-art results in terms of SSIM and PSNR for all the three datasets whose results are available. Qualitative results for indoor images indicate preservation of colors in the reconstructed image in the I-Haze dataset while a poorer color reconstruction is observed in the results of other datasets. In the future, we will improve our model to inherently include color preservation and seamless color cast as well. Source code, results, and trained model are shared at our project page ( https://github.com/ayu-22/BPPNet-Back-Projected-Pyramid-Network ).

\clearpage
%
%
\bibliographystyle{splncs04}
\bibliography{Bibliographies}
\end{document}